\documentclass{emulateapj} \usepackage{psfig} \usepackage{apjfonts}

\usepackage{graphicx}

\newcommand\beq{\begin{equation}}
\newcommand\eeq{\end{equation}}
\newcommand\lsim{\mathrel{\rlap{\lower4pt\hbox{\hskip1pt$\sim$}}
        \raise1pt\hbox{$<$}}}
\newcommand\gsim{\mathrel{\rlap{\lower4pt\hbox{\hskip1pt$\sim$}}
        \raise1pt\hbox{$>$}}}

\begin{document}

\title{A unified model of the magnetar and radio pulsar bursting phenomenology}

\author{Rosalba Perna\altaffilmark{1} and Jose A. Pons\altaffilmark{2}} 
\altaffiltext{1}{JILA and Department of Astrophysical and Planetary Science,
University of Colorado at Boulder, 440 UCB, Boulder, CO, 80304}
\altaffiltext{2}{Department de Fisica Aplicada, Universitat d'Alacant,
Ap. Correus 99, 03080, Alacant, Spain}

\begin{abstract}

Anomalous X-ray Pulsars (AXPs) and Soft Gamma-Ray Repeaters (SGRs) are
young neutron stars (NSs) characterized by high X-ray quiescent
luminosities, outbursts, and, in the case of SGRs, sporadic giant
flares.  They are believed to be powered by ultra-strong magnetic
fields (hence dubbed magnetars).  The diversity of their observed
behaviours is however not understood, and made even more puzzling by
the discovery of magnetar-like bursts from "low-field" pulsars.  Here
we perform long-term 2D simulations that follow the evolution of
magnetic stresses in the crust; these, together with recent
calculations of the breaking stress of the neutron star crust, allow
us to establish when {\em starquakes} occur.  For the first time, we
provide a quantitative estimate of the burst energetics, event rate,
and location on the neutron star surface, which bear a direct
relevance for the interpretation of the overall magnetar
phenomenology.  Typically, an ``SGR-like'' object tends to be more
active than an ``AXP-like'' object or a ``high-$B$ radio pulsar'', but
there is no fundamental separation among what constitutes the apparent
different classes. Among the key elements that create the variety of
observed phenomena, age is more important than a small variation in
magnetic field strength.  We find that outbursts can also be produced
in old, lower-field pulsars ($B\sim$~a few $\times 10^{12}$~G), but
those events are much less frequent than in young, high-field
magnetars.

\end{abstract}


\section{Introduction}

Among the isolated NSs, particularly distinctive is
a  class of sources characterized by long periods ($P\sim
2-11$~s) and high quiescent X-ray luminosities ($L_x\sim
10^{33}-10^{35}$~erg s$^{-1}$), generally larger than their entire reservoir
of rotational energy.  These sources, historically classified as AXPs and SGRs,
often  display  stochastic bursts of X-rays, releasing
energies $\sim 10^{39}-10^{41}$~erg,  and, in the case of SGRs,
sporadic though very  energetic $\gamma$-ray flares, with typical
energetics $\sim 10^{44}-10^{45}$~erg. 

The most successful model to explain both the high quiescent X-ray
luminosities, as well as the X-ray bursts and giant $\gamma$-ray
flares, is the {\em magnetar} model \citep{TD1995,TD2001},
in which SGRs and AXPs are believed to be endowed with large magnetic
fields, $B\sim 10^{14}-10^{15}$~G, possibly resulting from an active
dynamo at birth.  After a magnetar is born, the internal magnetic
field is subject to a continuous evolution through the processes of
Ohmic dissipation, ambipolar diffusion, and Hall drift.  In the crust,
magnetic stresses are generally balanced by elastic stresses.
However, as the internal field evolves, local magnetic stresses can
occasionally become too strong to be balanced by the elastic strength
of the crust, which hence breaks, and the extra stored
magnetic/elastic energy becomes available for powering the bursts and
flares.

Despite the success of the magnetar model in explaining some general
features of the triggering mechanism of bursts and flares, some major
questions have been left unanswered. In particular: what determines
the frequency of the bursts? Why do some objects display giant flares
(SGRs), while others (AXPs) do not?  How frequent are the different
phenomena?  Why do burst locations on the NS surface generally
correlate with the pulsar phase (maximum of the quiescent X-ray
lightcurve) for AXPs, while they do not for SGRs? (see e.g. \cite{KG2004}
for a review of these properties).

Moreover, the discovery of magnetar-like X-ray bursts from the young
pulsar PSR~J1846-0258 \citep{Gav2008}, with an inferred surface
dipolar magnetic field $B_p= 4.9\times 10^{13}$~G, lower than the
traditionally considered magnetar range, and, more recently, the
discovery of SGR~0418+5729 with an even lower $B_p<7.5\times
10^{12}$~G \citep{Rea2010}, well within the range of the rotation
powered pulsars which do not display any bursting behaviour, has
raised another obvious question: why some ``high-$B$'' pulsars (PSR
J1119-6127 and PSR J1814-1744)
do not display any discernable X-ray emission nor outburst \citep{Cam2000},
while at least one case of ``low-$B$'' NS does, if the
magnetic field is their driving force ?  It is clear that the dipolar magnetic
field alone is not sufficient to account for this variety of behaviours.   
A {\em unified} physical framework is still lacking, and it constitutes a major puzzle in neutron star
physics \citep{Kaspi2010}.

Here we present the first investigation that combines long-term 2D simulations
of the coupled magneto-thermal evolution of the NS with the study of the
breaking of the NS crust. Our study, by using realistic values of the breaking stress
of the NS crust, allows us to estimate burst energetics, recurrence
times, and surface distribution.  Our approach allows us to identify
some key elements of the magnetar phenomenology, and to shed light on
what creates the variety of the -- often puzzling -- observed
behaviour.

\section{Modeling the coupled magnetic-thermal evolution of NSs and crustal fractures}

We follow the evolution (in axial symmetry) of the magnetic field in a
magnetar crust with the numerical code described in \cite{PG2007}.  As
in previous works, we restrict our study to magnetic field
configurations confined to the crust.  We note that one of the main products
of our calculations, i.e. heating by current dissipation, has little relevance in the core
where the conductivity is high and the heat released is lost by
neutrino emission.  Also important for our work, the changes in
elastic stresses and their effect in the crust are of no relevance in
a liquid core.  The only issue that may affect our results is the
global change of magnetic field geometry that a fully coupled
crust-core evolution may give, but the way in which this could change our
picture is not clear.  This shortcoming is caused by the difficulties
(both theoretical and numerical) associated to the superconducting
state of protons in a NS core, which makes it difficult to consider
the full problem at present, although we hope to be able to do so in
future work.

The initial configurations of the magnetic field include both a toroidal and a poloidal component.
The temperature of the crust at different ages is given by the results of \cite{PMG2009}, where the
interplay between the thermal and magnetic field evolution of magnetars was studied.
For simplicity, and for numerical limitations, here we assume that the crust is isothermal. 
As the magnetic field evolves, the crust moves
through a series of equilibrium states in which its elastic stress $\sigma_b(r,\theta,t)$ 
balances the (time-dependent) magnetic stress $M_{ij}(r,\theta,t)$ in each direction.
Assuming that equilibrium is reached at a certain time, $t^{\rm eq}$, we can define
\begin{equation}
M^{\rm eq}_{ij}(r,\theta)=\frac{B_i(r,\theta,t^{\rm eq}) B_j(r,\theta,t^{\rm eq})}{4\pi}
\sim \sigma_b(r,\theta,t^{\rm eq}) \;,
\label{eq:equil}
\end{equation}
where $r$ and $\theta$ indicate the radial and poloidal coordinates, respectively.

Recent molecular dynamical simulations \citep{Chu2010} have provided a fit
for the maximum stress that a NS crust can sustain
\begin{equation}
\sigma_b^{\rm max} = \left(0.0195-\frac{1.27}{\Gamma-71}\right)\; n_i\,
\frac{Z^2 e^2}{a}\;,
\label{eq:sigma}
\end{equation}
where $\Gamma=Z^2e^2/aT$ is the Coulomb coupling parameter,
$a=[3/(4\pi n_i)]^{1/3}$ is the ion sphere radius, $n_i$ the ion
number density, $Z$ the charge number, $e$ the electron charge, and
$T$ the temperature.  

At some time $t_b$ during the evolution, any component $ij$ of the local
magnetic stresses can occasionally 
depart from the previous equilibrium condition at $t^{\rm eq}$
by an amount which is comparable to, or exceeds, the breaking stress
of the crust, $\sigma_b^{\rm max}(r,t_b)$ (angle-independent because
of our assumption of isothermality);
hence the crust fractures, yielding a starquake, and a new equilibrium state is reestablished.   

Since our simulations follow the evolution of ${\bf B}(r,\theta,t)$,
and hence $M_{i,j}(r,\theta,t)$, we can map the time-dependent regions
in the NS crust which are subject to fractures.  
A fully consistent dynamical simulation of the starquake is out of our present capabilities. The
typical timescales (mseconds to seconds) of individual burst/flare events are
many orders of magnitude smaller than the long-term evolution
timescales (typical timesteps in our code are 
$\sim$ a week, to allow us to follow the NS evolution up to
$10^5-10^6$ years), hence we cannot dynamically follow the fracture
propagation or model individual bursts. 
However, we can estimate the energy of an ``outburst"
(i.e. a collection of tens to hundreds of bursts occurring within 
$\sim$ a week timescale) as
follows. When a starquake happens, say at time $t=t_b$, a certain
region of the crust in the vicinity of the point where critical
conditions are reached will be affected. 
We consider that all surrounding regions where local 
magnetic stresses are close to the maximum
(say by a fraction $\epsilon$) are affected, i.e.
\begin{equation}
M_{ij}(r,\theta,t_b) - M^{\rm eq}_{ij}(r,\theta) \ge \epsilon \sigma_b^{\rm max}(r,t_b) \;.
\label{eq:break2}
\end{equation}

This parameter $\epsilon$ must be close to the fatigue limit
of the material (for terrestrial materials it ranges between 35\% and 60\%).
This limit is also close to the yield point:
the stress at which a material begins to deform plastically. 
All investigated crystallographic shear systems in 
\cite{HK2009} break in a rather abrupt fashion
with only a small region where plasticity is present.
Thus, we expect the parameter $\epsilon$ to be in the range [0.8-0.99].
It parametrizes our ignorance about the transition from elastic
to plastic regimes and the fatigue of the material.

We can compute the elastic energy stored in that portion of the crust,
and assume that it will be released at once (one timestep)
and that the affected region will return immediately to equilibrium.
The new equilibrium stresses are reset, and the process is repeated.
We neglect the feedback produced by the
local deposition of energy, which will require a much more complex
modeling.  In this way, we estimate the energy available in a given
event, the time interval between events, and the location of the
fractures. It should be stressed that this represents the total energy
of an ``outburst", that may be released in one very energetic flare,
many small bursts, or a combination of both.  Also, depending on
the local physical conditions, part of the
energy is lost to neutrinos, part is transferred to the magnetosphere,
and part results in local heating and is radiated by photons from the
surface on a much longer timescale. Typically, energy released in the
inner crust is more easily lost to  neutrinos while energy
released near the surface has a more direct
observational impact through thermal (surface) and non-thermal
(magnetospheric) radiation.

The energy available after a starquake  can therefore be estimated as the elastic energy
corresponding to a static shear strain $\Psi$, which can be well approximated by \citep{TD2001}
\begin{equation}
E_b(t_b) = \frac{1}{2} \int dV  \mu \Psi^2   = \Sigma_{i,j}  \int dV  \frac{
\left[M_{ij}(r,\theta,t_b) -M^{\rm eq}_{ij}(r,\theta)\right]^2}{\mu(r,t_b)}
\label{eq:energy}
\end{equation}
where we have used the fact that 
the yield strain is approximated by
\begin{equation}
\Psi_{ij}  \approx  \frac{\left[M_{ij}(r,\theta,t_b) -M^{\rm eq}_{ij}(r,\theta)\right]}{\mu(r,t_b)},
\label{eq:energy2}
\end{equation}
and the shear modulus $\mu$ is given by \citep{HH2008}
\begin{equation}
\mu(r,t) = \left(0.1106-\frac{28.7}{\Gamma^{1.3}}\right)\; n_i\,
\frac{Z^2 e^2}{a}\,.
\label{eq:mu}
\end{equation}
The integral in Eq. (\ref{eq:energy}) is computed over the volume $dV$
for which condition (\ref{eq:break2}) on the stresses in any direction
is satisfied. 
Additional energy is stored in the magnetic field, 
but the fraction of this energy released during a fracture is
unknown, and depends on the amount of slippage. This 
requires a proper calculation involving fracture dynamics on short
timescales. To be conservative, we consider the case in which only
elastic energy is released, so our energy estimates are lower limits.

\section{Predicting starquake energetics, frequencies and distribution on NS surface}

We now report the most relevant results. As a baseline model we have chosen a NS
born with an initial dipolar field of $B_p=8 \times 10^{14}$~G
and an internal toroidal field of $B_t=2 \times
10^{15}$~G at maximum, similar to models used in previous studies and close to what one expects to be
the final geometry after MHD equilibrium is reached in a hot, liquid, proto-NS
\citep{Ciolfi2009,Ciolfi2010,LJ2009}.  In this
particular model, at the age of $10^4$ yrs the dipolar field has
decreased to about 20\%  of its initial value, while its internal toroidal
field has been dissipated by more than a factor of 2.  Other
initial conditions modify quantitatively the results, but the general trends remain nearly the same. The
evolution is followed for $10^5$~years.  During this time, we monitor
the frequency, angular distribution, and energetics of the fractures.

\begin{figure}[ht!]
\includegraphics[width=7.9cm]{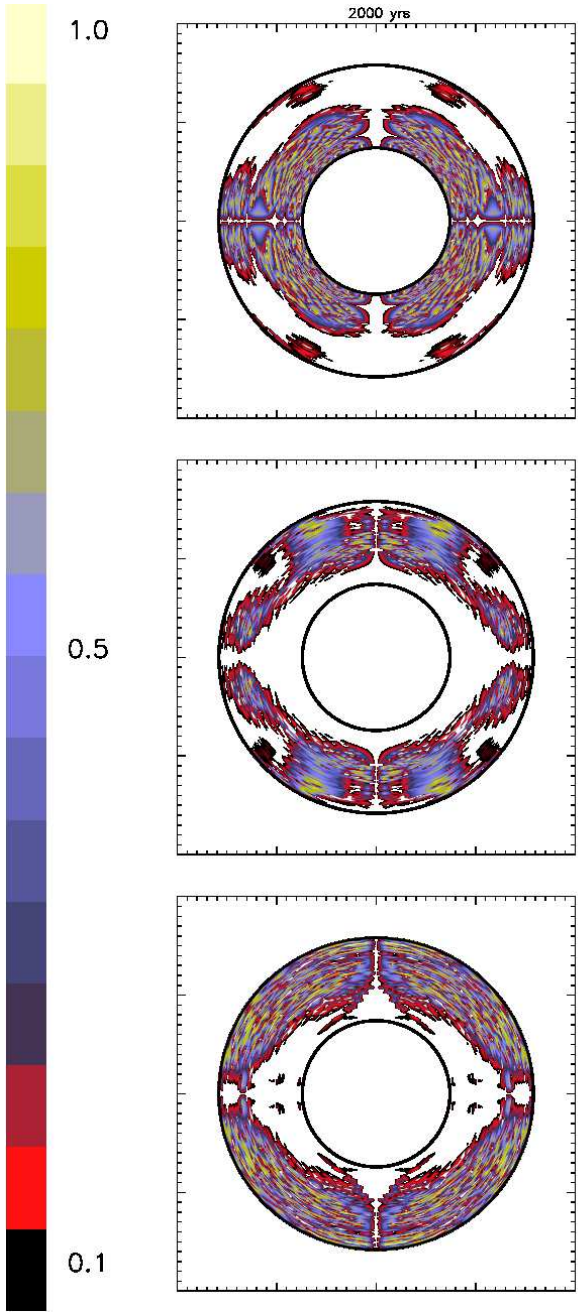}
\caption{\small Snapshot of the deviation of stresses from the
previous equilibrium in a typical 2000 yr old magnetar crust. 
The top, middle and bottom panels show the $(\theta \phi)$, $(r \phi)$,
and $(r \theta)$ components of the tensor $M_{ij}-M^{\rm eq}_{ij}$
normalized to the local value of $\sigma_b^{\rm max}$. The
crustal region has been stretched by a factor of 6 for clarity of
visualization. The components of the stress tensor are normalized to
the local value of the breaking stress, so that yellow regions are
close to the fracture limit.}
\end{figure}

Fig.~1 shows a snapshot of the deviation of stresses from the
previous equilibrium in a typical 2000 yr old magnetar.   The yellow
regions are close to the breaking limit. Note that at this age the
crust has already suffered hundreds of fractures.
That history results in the patched appearance.
For this run we have fixed $\epsilon = 0.9$.

The results about frequency and energy distribution of events are contained in
Fig.~2.  We remind again that groups of bursts connected to the same
fracture are classified as a single outburst in our simulations.  We
selected three representative periods in a magnetar life, each
involving the same total number of events (1000).  The first is
labeled "Young" and spans the interval between 400 and 1600 yrs. The
second is labeled "Mid age" and covers the period $\approx 7-10$ kyrs, and
the third period is named "Old" and corresponds to the events
recorded from 60 to 100 kyrs.  
Note that this classification does not pretend to
reflect an exact correspondence to SGRs and AXPs, which is just terminology based on historical
reasons. 
Objects that could belong to the first category (young or "SGR-like") are
SGR 1806-20,  SGR 1900+14 and 1E 1547.0-5408; to the Mid Age or "AXP-like" list
could belong SGR 1627-41, SGR 0526-66, SGR 0501+4516,  1E1048.1-5937, or 1E 2259+586
among others; finally, examples of old objects could be 
SGR 0418+5729, SGR 1833-0832,  XTE J1810-197, and PSR~J1814-1744.

The first important result is that there is a significant difference
in the energetics and recurrence time as the star evolves. This is
because the initial evolution of the crustal magnetic field is faster due
to the Hall drift, while at late times its has been rearranged
into a more quasi-steady state and the evolution proceeds slower. In young
magnetars, the typical energies released after each starquake are of
the order of $10^{44}$~erg and the typical event rate $\sim$
1/yr. This does not necessarily imply that a flare will be
observed, since this is the energy released
in the crust and the mechanism to transfer energy to the
surface and magnetosphere may not work in many cases, especially when
the fracture occurs in the inner crust. However, in our simulations we
observed that more than 95\% of the fractures occur in the outer crust,
simply because the crust is less strong there, so that we
expect a relatively high efficiency of the process.  For mid-age
objects the energetics is shifted to lower values, a second peak
appears at about $10^{41}$~erg, and the waiting time between outbursts
increases to a few years. This trend is more pronounced as the object
gets older: the recurrence time becomes of the
order of tens of years and nearly half the events are low-energy
events.  The physical reason behind the bimodal distribution is the
directionality of stresses. We found that fractures associated to the
magnetic stress component $M_{\theta,\phi}$ are more frequent, but
they are mostly associated to very low energy events $\lsim
10^{41}$~erg. On the other hand, fractures caused by the $M_{r,\phi}$
component are responsible for most of the $E\gsim10^{44}$~erg
events. The events associated to large values of $M_{r,\theta}$ span
the whole range of energies, but they become very rare after a few kyrs,
so that the long-term energy distribution becomes
more clearly bimodal. On average, we find that events related to
stresses created by the toroidal field are a hundred times more
frequent.  Our simulations hence predict that giant flares are
expected to be less frequent than the less energetic bursts.  Indeed,
while giant flares have been observed in only 3 objects (once each),
bursts have been observed in more than a dozen (often multiple times
each).

In Fig. 3 we show the results for the same case, but with the
parameter $\epsilon$ varying randomly in the interval [0.95-1] for
each event. There is no reason why this parameter must be constant and
in fact fracture dynamics cannot be
parametrized in a simple form. Nevertheless, with this other study, we
can assess the sensitivity of our results to the particular choice of
$\epsilon$. The first important remark is that the energy distribution
changes and now low-energy events are more frequent than
giant-flare-like events even for young magnetars. This choice of values of $\epsilon$ closer to unity
reduces the average size of the region implied in the starquake, and selects less energetic events.
Interestingly, the bimodal shape of the energy distribution remains qualitatively the
same. In the central panel we can see
that the age trend discussed in the previous case does not change: the
older the object, the longer the average waiting times between two
events. The comparison between these two cases indicates that a
detailed understanding and modeling of fracture mechanics will be
needed before we can make definite, quantitative statements about the
energy distribution of starquakes.

\begin{figure*}[ht!]
\includegraphics[width=18cm]{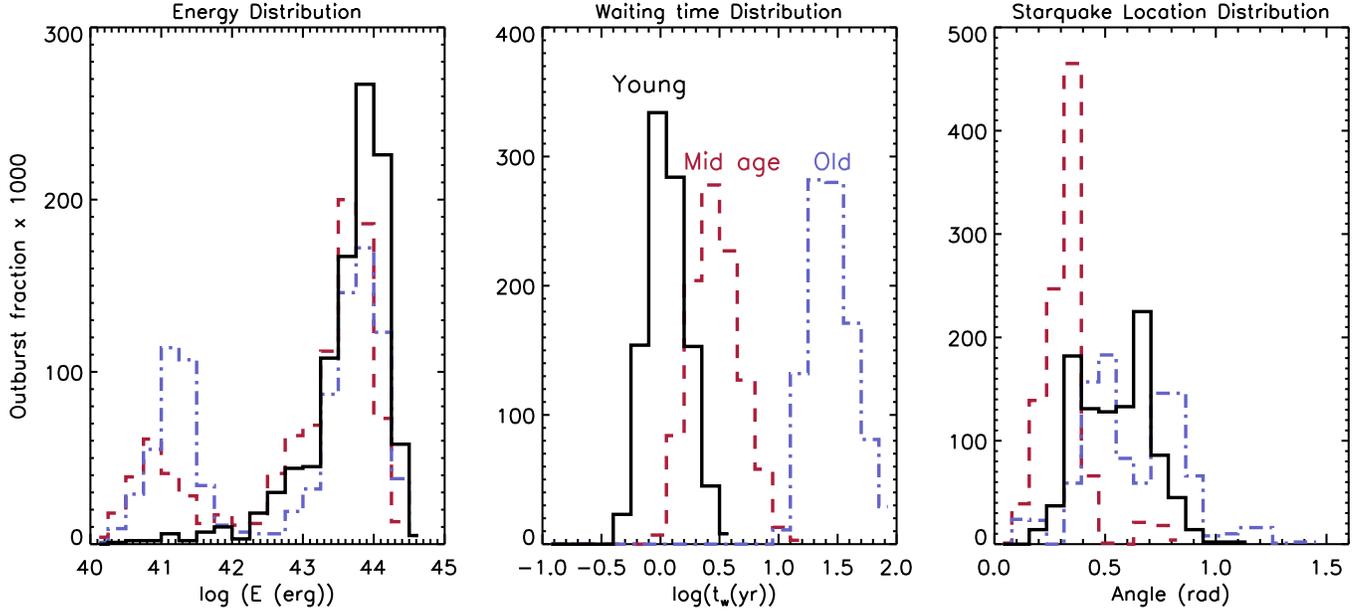}
\caption{\small Outburst properties of an object with initial magnetic field
components $B_p=8\times 10^{14}$~G and $B_t=2\times 10^{15}$~G during
three different periods of its lifetime: 400-1600~yr (labeled ``Young''), 7-10~kyr
(labeled as ``Mid age''), and 60-100~kry (labeled as ``Old''). We have fixed $\epsilon=0.9$.}
\end{figure*}

\begin{figure*}[ht!]
\includegraphics[width=18cm]{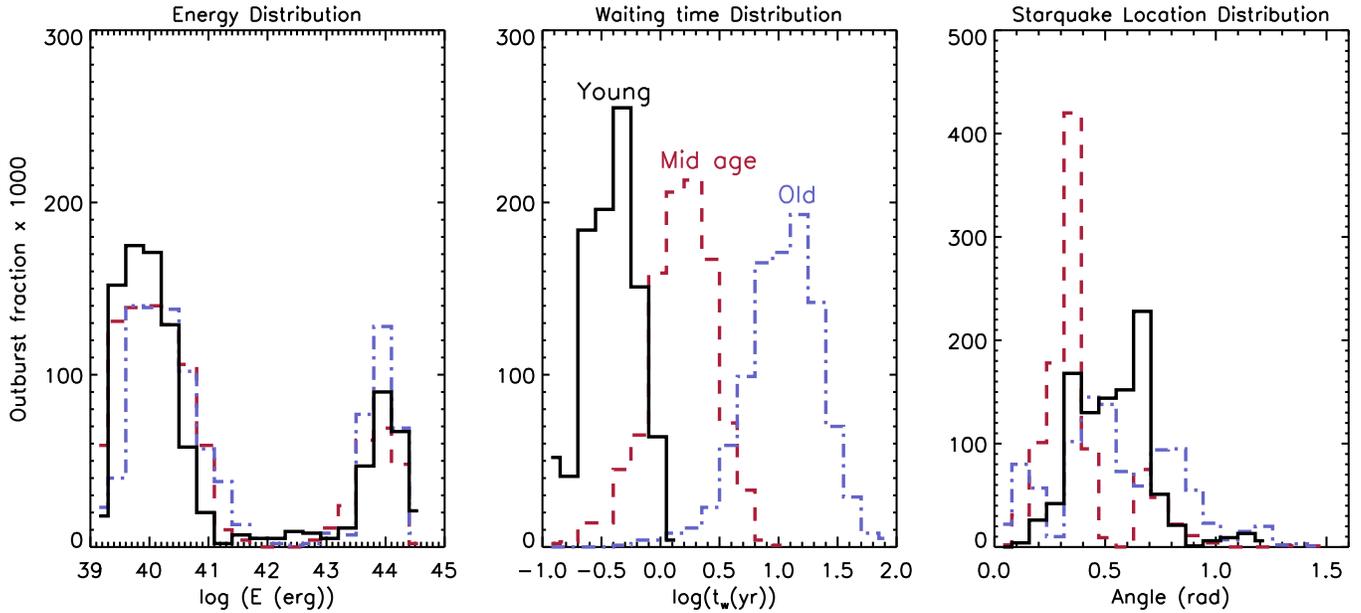}
\caption{\small Same as Fig. 2, but with $\epsilon$ varying randomly in the
interval [0.95-1].}
\end{figure*}

It is interesting to compare the general trends predicted by
our simulations with the observations. We find
that, for a given $B$-field strength and configuration at birth,
younger objects are generally more active, both in event frequency and
energy of events. Interestingly, this
age sequence is also what observations have been hinting at
\citep{Kou1999}.  Our predicted clustering of the energy
distribution (around $10^{40}-10^{41}$ and $10^{44}$~erg) is
suggestive of the observed dichotomy outbursts/giant flares, with 
rarer events in between \citep{Kou2001,Israel2008,Mere2008}. 
Note that our results predict that middle age objects also show high energy
events; interestingly, some of the originally classified AXPs also show SGR-like bursts \citep{KS2010}.
A rigorous comparison with observations cannot be
made at this stage, due to the very limited statistics. However, if
the secular waiting time distribution of starquakes (which is what we
predict) has a similar behaviour to the waiting time distribution of
bursts within an outburst, then our results would again match the
observations \citep{Cheng1996,Gogus2001}.

Further connections to the observations can be made by inspection of
the right panels of Figs. 2 and 3, where we show the angular
distribution of the bursts. While for "young" and "old" magnetars there is
no clear trend, it is interesting to note that at intermediate age 
most of the fractures happen at small polar angles. If the
polar region is hotter than the equator, our results suggest a
correlation between burst location and pulsar phase, which is 
typical of objects routinely classified as AXPs
\citep{KG2004,Gav2004}.  The reason for this
particular feature is probably related to the evolution
of the internal toroidal field. The Hall drift displaces the internal
toroidal field towards the poles in a typical Hall timescale of
$10^3-10^4$ yrs, but after a few Hall timescales the magnetic field is
reconfigured in a more stable, steady-state. The
characteristic location of starquakes closer to the pole seems to be reflecting
this fact. To confirm this point, consistent simulations that include the local heating and the
temperature variations produced by the energy release are in progress
and will be presented elsewhere.

Finally, we also studied the magnetic evolution of a NS with a lower
dipolar magnetic field strength $B_p=2\times 10^{14}$~G, and a
toroidal component $B_t=10^{15}$~G (at birth).  We made a longer simulation covering 
$10^6$~yr of this NS and found that, while the energetics is very similar,
the event rate is much lower. Even at very early times the typical
lapse time between events is about 50-100 yrs. At late times, it
becomes larger, ranging from few kyrs to 10 kyrs. The reason why the
energetics is similar is simple: the local physical conditions that
establish the breaking criteria are the same (Eq. \ref{eq:break2}), so the
elastic energy stored is similar. However, the magnetic field evolves
slowly, so it takes more time to bring the system close to the
critical conditions. Interestingly, our magneto-thermal evolution
model naturally predicts that {\em all NSs, not only magnetars, may
show activity in the form of sporadic outbursts or flares}. This last
case, for which $B_p$ becomes $< 7\times 10^{12}$~G at an age $> 1$~ Myr,  
would be a close representation of the bursting source recently
reported by \cite{Rea2010}.
Other cases studied show similar results, with the particularity that
objects with a lower toroidal field are generally less active.

\section{Summary}

In our study, we have found that there is no real separation among
what constitutes the apparently different classes of bursting NSs,
and we have identified some key elements that create the
variety of observed burst/flare phenomenology.  In particular, while what we infer
from measurements of $P$ and $\dot{P}$ is only the dipolar component
of the $B$-field, both the dipolar component and the hidden toroidal
component are similarly important.  The lower the field (in either
component), the lower is the frequency of starquakes.  Although very
rare, we find that outbursts can also occur in ``low-$B$'' pulsars,
with $B\sim$ a few $\times 10^{12}$~G.  For a given $B$ field at birth,
our results show that the age (or the evolutionary stage) is the most
relevant feature to determine the different levels of activity in
different families of NSs. The same object may behave differently at
different times, without in principle having a different nature.

\acknowledgements
This work has been supported in part by NASA grants NNX10AK78G and GO9-07710X (RP)
and by the Spanish MEC grant AYA 2010-21097-C03-02, GVPROMETE02009-103, 
and the Research Network Program Compstar funded by the ESF (JP).
We thank S. Mereghetti, N. Rea and an anonymous referee for useful
comments on the manuscript.

\end{document}